\documentclass[10pt]{IEEEtran} 
\usepackage{graphicx}
\usepackage{epstopdf}
\usepackage[usenames,dvipsnames]{xcolor}
\usepackage{wrapfig}
\usepackage{setspace}
\usepackage{cite}
\usepackage[hidelinks]{hyperref}
\usepackage{amsmath}
\usepackage{diagbox}
\usepackage{multirow}
\usepackage{todonotes}
\usepackage{color}
\usepackage{algorithm}
\usepackage[noend]{algpseudocode}

\begin{document}

\title{Dynamic Modeling of Cascading Failure in Power Systems}
\author{Jiajia Song, \IEEEmembership{Graduate Student Member, IEEE,} Eduardo Cotilla-Sanchez, \IEEEmembership{Member, IEEE}\\Goodarz Ghanavati, \IEEEmembership{Graduate Student Member, IEEE,} Paul D.~H.~Hines, \IEEEmembership{Senior Member, IEEE} 
\thanks{Manuscript received November 14, 2014. This work was supported in part by the U.S.~Department of Energy.}
\thanks{J.~Song and E.~Cotilla-Sanchez are with the School of Electrical Engineering \& Computer Science at Oregon State University, Corvallis, OR, 97331, USA (e-mail:~\{songjia,ecs\}@eecs.oregonstate.edu). 

G.~Ghanavati and P.~Hines are with the School of Engineering at the University of Vermont, Burlington, VT 05405, USA (e-mail:~\{gghanava,paul.hines\}@uvm.edu)}}

\IEEEpubid{0000--0000/00\$00.00?\copyright?2014 IEEE }%

\maketitle

\bstctlcite{IEEEexample:BSTcontrol}

\begin{abstract}The modeling of cascading failure in power systems is difficult because of the many different mechanisms involved; no single model captures all of these mechanisms. Understanding the relative importance of these different mechanisms is important for choosing which mechanisms need to be modeled for particular applications. This work presents a dynamic simulation model of both power networks and protection systems, which can simulate a wider variety of cascading outage mechanisms relative to existing quasi-steady state (QSS) models. This paper describes the model and demonstrates how different mechanisms interact. In order to test the model we simulated a batch of randomly selected $N-2$ contingencies for several different static load configurations, and found that the distributions of blackout sizes and event lengths from the simulator correlate well with historical trends. The results also show that load models have significant impacts on the cascading risks. Finally, the dynamic model was compared against a simple dc-power-flow based QSS model; we find that the two models tend to agree for the early stages of cascading, but produce substantially different results for later stages. 

\end{abstract}

\begin{IEEEkeywords} Cascading outages, cascading failures, power system dynamic simulation, differential algebraic equation, power system modeling, power system protection.
\end{IEEEkeywords}

\section{Introduction} \label{intro}
\IEEEPARstart{T}{he} vital significance of studying cascading outages has been recognized in both the power industry and academia \cite{Baldick:2008,Papic:2011,Vaiman:2012}. However, since electrical power networks are very large and complex systems \cite{Dobson:2007}, understanding the many mechanisms by which cascading outages propagate is challenging. This paper presents the design of and results from a new non-linear dynamic model of cascading failure in power systems (the Cascading Outage Simulator with Multiprocess Integration Capabilities or COSMIC), which can be used to study a wide variety of different mechanisms of cascading outages.

A variety of cascading failure modeling approaches have been reported in the research literature, many of which are reviewed in \cite{Papic:2011,Baldick:2008,Vaiman:2012}. Several have used quasi-steady state (QSS) dc power flow models \cite{Eppstein:2012,Carreras:2002,Yan:2014}, which are numerically robust and can describe cascading overloads. However, they do not capture non-linear mechanisms like voltage collapse or dynamic instability. QSS ac power flow models have been used to model cascading failures in \cite{Nedic:2006,Chen:2013,Mei:2008,Bienstock:2011}, but these models still require difficult assumptions to model machine dynamics and to deal with non-convergent power flows. Some have proposed models that combine the dc approximations and dynamic models \cite{Fitzmaurice:2012}, allowing for more accurate modeling of under-voltage and under-frequency load shedding. These methods increase the modeling fidelity over pure dc models but still neglect voltage collapse. Others developed statistical models that use data from simulations \cite{Rahnamay-Naeini:2014,Hines:2013} or historical cascades \cite{Dobson:2012} to represent the general features of cascading. Statistical models are useful, but cannot replace detailed simulations to understand particular cascading mechanisms in depth. There are also topological models \cite{Boccaletti:2006,Hines:2011,Wang:2009,Albert:2004} which have been applied to the identification of vulnerable/critical elements, however, without detailed power grid information, the results that they yield differ greatly and could result in misleading conclusions about the grid vulnerability \cite{Hines:2010}. 

Some dynamic models \cite{Ma:2007,Fabozzi:2009} and numerical techniques \cite{Khaitan:2009,Abhyankar:2012} study the mid/long-term dynamics of power system behavior, and show that mid/long term stability is an important part of cascading outage mechanisms. However, concurrent modeling of power system dynamics and discrete protection events---such as line tripping by over-current, distance and temperature relays, under-voltage and under-frequency load shedding---is challenging and not considered in most existing models. In \cite{Parmer:2011} the authors describe an initial approach using a system of differential-algebraic equations with an additional set of discrete equations to dynamically model cascading failures. 

\IEEEpubidadjcol
The paper describes the details of and results from a new non-linear dynamic model of cascading failure in power systems, which we call ``Cascading Outage Simulator with Multiprocess Integration Capabilities'' (COSMIC). In COSMIC, dynamic components, such as rotating machines, exciters, and governors, are modeled using differential equations. The associated power flows are represented using non-linear power flow equations. Load voltage responses are explicitly represented, and discrete changes (e.g., components failures, load shedding) are described by a set of equations that indicate the proximity to thresholds that trigger discrete changes. Given dynamic data for a power system and a set of exogenous disturbances that may trigger a cascade, COSMIC uses a recursive process to compute the impact of the triggering event(s) by solving the differential-algebraic equations (DAEs) while monitoring for discrete events, including events that subdivide the network into islands.

There are only a few existing commercial and research-grade simulation tools that specifically address cascading failure events and their consequences \cite{Papic:2011}. Static modeling is still dominant in these tools, although at least two commercial packages (Eurostag ASSESS \cite{Antoine:1992} and POM-PCM \cite{Bhatt:2009}) have introduced dynamic simulation to their cascading failure analysis. The model that we propose here supplements these commercial tools by providing an open platform for research and development that allows one to explicitly test the impact of the many assumptions that are necessary for dynamic cascading failure modeling. For example, users can modify the existing system components, add new ones, and integrate advanced remedial control actions. Additionally, the dynamic/adaptive time step and recursive islanded time horizons  (see Section \ref{solve_dae}) implemented in this simulator allow for faster computations during, or near, steady-state regimes, and fine resolution during transient phases. Moreover, this tool can be easily integrated with High Performance Computing (HPC) clusters to run many simulations simultaneously at a much lower cost, relative to commercial tools.

The remainder of this paper proceeds as follows: Section \ref{sec:dae} introduces the components of the model mathematically and describes how different modules interact. In Section \ref{sec:results}, we present results from several experimental validation studies. Section \ref{sec:conclusion} presents our conclusions from this study. Finally, Section \ref{sec:appendix} provides further details on the model and its settings. 

\section{Hybrid system modeling in COSMIC} \label{sec:dae}
\subsection{Hybrid differential-algebraic formulation}
Dynamic power networks are typically modeled as sets of DAEs \cite{Kundur:1994}. If one also considers the dynamics resulting from discrete changes such as those caused by protective relays, an additional set of discrete equations is added, which results in a hybrid DAE system \cite{Van-Der-Schaft:2000}.

Let us assume that the state of the power system at time $t$ can be defined by three vectors: $\mathbf{x}(t)$, $\mathbf{y}(t)$, and $\mathbf{z}(t)$, where:
\begin{description}
\item [{$\mathbf{x}$}] is a vector of continuous state variables that change with time according to a set of differential equations
\begin{equation}
\label{eqs:diff_equation}
\frac{d\mathbf{x}}{dt}=\mathbf{f}(t,\mathbf{x}(t),\mathbf{y}(t),\mathbf{z}(t))
\end{equation}
\item [{$\mathbf{y}$}] is a vector of continuous state variables that have pure algebraic relationships to other variables in the system:
\begin{equation}
\label{eqs:alg_equation}
\mathbf{g}(t,\mathbf{x}(t),\mathbf{y}(t),\mathbf{z}(t))=0
\end{equation}
\item [{$\mathbf{z}$}] is a vector of state variables that can only take integer states ($z_{i}\in[0,1]$)
\begin{equation}
\label{eqs:discret_equation}
\mathbf{h}(t,\mathbf{x}(t),\mathbf{y}(t),\mathbf{z}(t))<0
\end{equation}
\end{description}
When constraint $\mathbf{h}_i(...)<0$ fails, an associated counter function $\mathbf{d}_i$ (see \ref{relay_modeling}) activates. Each $\mathbf{z}_i$ changes state if $\mathbf{d}_i$ reaches its limit. 

The set of differential equations \eqref{eqs:diff_equation} represent the machine dynamics (and/or load dynamics if dynamic load models are included). In COSMIC the differential equations include a third order machine model and somewhat simplified governor and exciter models in order to improve computational efficiency without compromising the fundamental functions of those components (see Section \ref{app_A} for more details). In particular, the governor is rate and rail limited to model the practical constraints of generator power control systems. The governor model incorporates both droop control and integral control, which is important to mid/long term stability modeling, especially in isolated systems \cite{Kundur:1994}. 

The algebraic constraints \eqref{eqs:alg_equation} encapsulate the standard ac power flow equations. In this study we implemented both polar and rectangular power flow formulations. Load models are an important part of the algebraic equations, which are particularly critical components of cascading failure simulation because i) they need to represent the aggregated dynamics of many complicated devices and ii) they can dramatically change system dynamics. The baseline load model in COSMIC is a static model, which can be configured as constant power ($P$), constant current ($I$), constant impedance ($Z$), exponential ($E$), or any combination thereof ($ZIPE$) \cite{Song:2013}. 

\begin{figure}[h]
\centering
\includegraphics[width=1\columnwidth]{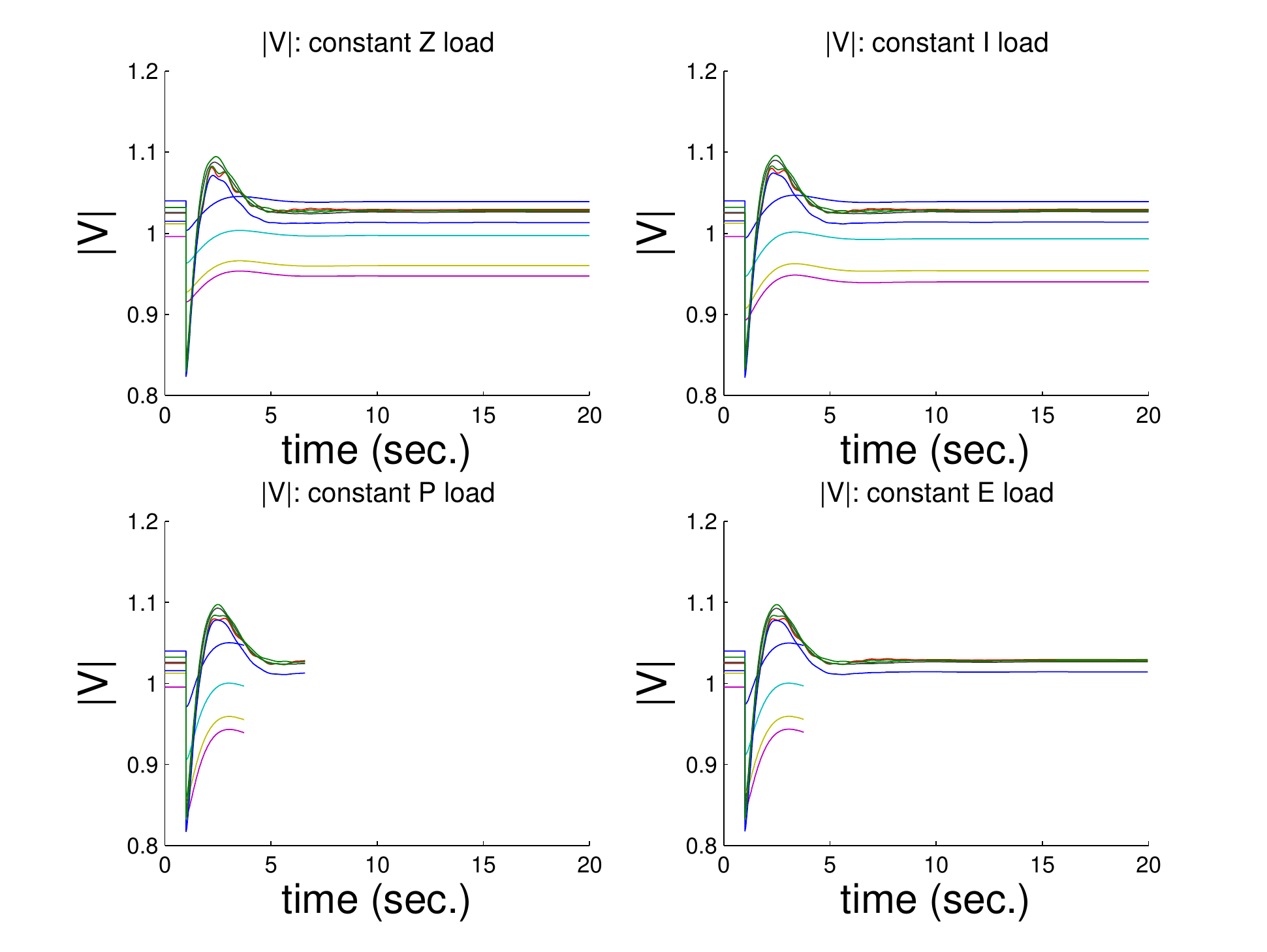}
\caption{Bus voltage responses to two line outages (5-7 and 6-9) in the IEEE 9-bus for the baseline load types in the COSMIC model. The system separates into two islands after the initial events ($t=1$ second). The non-converged network is declared as blacked-out area in which all the algebraic and differential variables are not available anymore (see bottom left and right panels for the P and E load models as examples of this behavior).}
\label{Fig:V_zipe}
\end{figure}

As Fig.~\ref{Fig:V_zipe} illustrates, load models can have a dramatic impact on algebraic convergence. Constant power loads are particularly difficult to model for the off-nominal condition. Numerical failures are much less common with constant I or Z loads, but are not accurate representations of many loads. This motivated us to include the exponential component in COSMIC.

During cascading failures, power systems undergo many discrete changes that are caused by exogenous events (e.g., manual operations, weather) and endogenous events (e.g., automatic protective relay actions). The discrete event(s) will consequently change algebraic equations and the systems dynamic response, which may result in cascading failures, system islanding, and large blackouts. In COSMIC, the endogenous responses of a power network to stresses are represented by \eqref{eqs:discret_equation}. These discrete responses are described in detail in \ref{relay_modeling} and \ref{solve_dae}.

\subsection{Relay modeling} \label{relay_modeling}
Major disturbances cause system oscillations as the system seeks a new equilibrium. These oscillations may naturally die out due to the interactions of system inertia, damping, and exciter and governor controls. In order to ensure that relays do not trip due to brief transient state changes, time-delays are added to each protective relay in COSMIC.

We implemented in this model two types of time-delayed triggering algorithms: fixed-time delay and time-inverse delay. These two delay algorithms are modeled by a counter function, $\mathbf{d}$,  which is triggered by \eqref{eqs:discret_equation}. The fixed-time delay triggering activates its counter/timer as soon as the monitored signal exceeds its threshold. If the signal remains beyond the threshold, this timer will continue to count down from a preset value until it runs out then the associated relay takes actions. Similarly, the timer will recover if the signal is within the threshold and will max out at the preset value. For the time-inverse delay algorithm, instead of counting for the increment of time beyond (or within) the threshold, we evaluate the area over (or under) the threshold by integration (based on Euler's rule).

Five types of protective relays are modeled in COSMIC: over-current (OC) relays, distance (DIST) relays, and temperature (TEMP) relays for transmission line protection; as well as under-voltage load shedding (UVLS) and under-frequency load shedding (UFLS) relays for stress mitigation. OC relays monitor the instantaneous current flow along each branch. DIST relays represent a Zone 1 relay that monitors the apparent admittance of the transmission line. TEMP relays monitor the line temperature, which is obtained from a first order differential equation
\begin{equation}
\dot{T_{i}}=r_{i}F^2_{i}-k_{i}T_{i}
\end{equation}
where $T_{i}$ is the temperature difference relative to the ambient temperature (20 $^{\circ}$C) for line $i$, $F_{i}$ is the current flow of line $i$, $r_{i}$ and $k_{i}$ are the are heating and time constants for line $i$ \cite{Fitzmaurice:2012}. $r_{i}$ and $k_{i}$ are chosen so that line $i$'s temperature reaches 75 $^{\circ}$C (ACSR conductors) if current flow hits the rate-A limit, and its TEMP relay triggers in 60 seconds when current flow jumps from rate-A to rate-C. The threshold for each TEMP relay is obtained from the rate-B limit. While it would have been possible to incorporate the temperature relays into the trapezoidal integration used for the other differential equations, these variables are much slower than the other differential variables in $\mathbf{x}$. Instead, we computed $T_{i}$ outside of the primary integration system using Euler's rule. 

When voltage magnitude or frequency signals at load Bus $i$ are lower than the specified thresholds, the UVLS relay or UFLS relay will shed a $25\%$ (default setting) of the initial $P_{d,i}$ to avoid the onset of voltage instability and reduce system stress. In order to monitor frequency at each load bus, Dijkstra's algorithm \cite{Dijkstra_algorithm} and electrical distances \cite{Cotilla-Sanchez:2012} were used to find the generator (and thus frequency from $\mathbf{x}$) that is most proximate to each load bus. Both the UVLS and UFLS relays used a fixed-time delay of $0.5$ seconds.

\subsection{Solving the hybrid DAE} \label{solve_dae}
Because of its numerical stability advantages, COSMIC uses the trapezoidal rule \cite{Milano:2010} to simultaneously integrate and solve the differential and algebraic equations (see Section \ref{app_B} for more details).

Whereas many of the common tools in the literature \cite{Siemens:2011,powerworld} use a fixed time step-size, COSMIC implements a variable time step-size in order to trade-off between the diverse time-scales of the dynamics that we implement. We select small step sizes during transition periods that have high deviation or oscillations in order to keep the numerical within tolerance; the step sizes increase as the oscillations dampen toward steady-state values. 

When a discrete event occurs at $t_{d}$, the complete representation of the system is provided in the following equations:
\begin{align}
0 & = &\mathbf{x}+\frac{t_{d}-t}{2}\left[\mathbf{f}(t)+\mathbf{f}(t_{d},\mathbf{x}_{d},\mathbf{y}_{d},\mathbf{z})\right] \label{eq:discrete1}
\\ 0 & = & \mathbf{g}(t_{d},\mathbf{x}_{+},\mathbf{y}_{+},\mathbf{z}) \label{eq:discrete2}
\\ 0 & > & \mathbf{h}(t_{d},\mathbf{x}_{+},\mathbf{y}_{+})\label{eq:discrete3}
\\ 0 &=& \mathbf{d}(t_{d},\mathbf{x}_{+},\mathbf{y}_{+}) \label{eq:delay}
\end{align}
where, $t$ is the previous time point, and $\mathbf{d}$ is the counter function mentioned previously. Because of the adaptive step-size, COSMIC retains $t_{d}$ from $t_{d} = t+\Delta t_{d}$, in which $\Delta t_{d}$ is found by linear interpolation of two time steps. 

Every time a discrete event happens ($\mathbf{h}<0$ and $\mathbf{d}=0$), COSMIC stops solving the DAE for the previous network configuration, processes the discrete event(s), then resumes the DAE solver using the updated initial condition. Specifically, it uses updated algebraic variables while holding the differential variables. In some cases the discrete event causes the system to split into multiple islands. COSMIC checks for this condition by inspection of the network admittance matrix ($Y_{\text{bus}}$) (see Algorithm 1). COSMIC deals with the separation of a power network into sub-networks (unintentional islanding) using a recursive process that is described in Algorithm 1. If islanding results from a discrete event, the present hybrid DAE separates into two sets of DAEs. 
COSMIC treats the two sub-networks the same way as the original one, integrates and solves these two DAE systems in parallel, and synchronizes two result sets in the end. 

One can find the relevant parametric settings for the numerical integration and protective relays in Table \ref{table_num_para} and Table \ref{table_relay_para} (Section \ref{app_C}).

\makeatletter
\def\BState{\State\hskip-\ALG@thistlm}
\makeatother
\begin{algorithm}

\caption{Time-Domain Simulation Algorithm} 
\begin{algorithmic}
\Procedure{}{}
\BState \textbf{Step 1}: Build and initialize the hybrid DAE system for the given power network.
\BState \textbf{Step 2}: Process the exogenous contingencies if they exist.
\BState \textbf{Step 3}: Check for network separation by inspecting the updated $Y_{\text{bus}}$. 
\If {Yes,}
\State Divide the current system into two sub-networks.
\State Recursively launch a time-domain simulation for each of the two sub-networks, starting at \textbf{Step 1}.
\EndIf

\BState \textbf{Step 4}: 
\If {the network configuration changed,} 
\State Recompute the $Y_{\text{bus}}$; re-solve the algebraic system $g(x,y,z)=0$ to find the new $y$.
\EndIf

\BState \textbf{Step 5}: Integrate the continuous DAE system until one of two conditions occur: (\textbf{a}) the simulation reaches its pre-specified end time ($t=t_{\text{max}}$), or (\textbf{b}) one of the discrete thresholds is crossed ($\mathbf{h}<0$).
\BState \textbf{Step 6}: Check for discrete events (Condition (\textbf{b}) in \textbf{Step 5}). 
\If {Yes,}
\State Identify the time point at which the discrete event(s) occurred ($h_i\le0$ and $\mathbf{d_i}=0$).
\State Find the values for differential variables, $x$, and algebraic variables, $y$, at the above time point using linear interpolation between time steps.
\State Process the discrete events by changing relay status ($z_i$) according to $h_i$.
\State Go back to \textbf{Step 3}.
\Else
\State Go back to \textbf{Step 5}.
\EndIf

\BState \textbf{Step 7}: Merge the time-series data from this sub-graph upstream with the rest of the system until reaching the top-level and then save the aggregate.
\EndProcedure
\end{algorithmic}
\end{algorithm}


\subsection{Validation} 
To validate COSMIC, we compared the dynamic response in COSMIC against commercial software---PowerWorld \cite{powerworld}---using the classic 9-bus test case \cite{Anderson:2008}. From a random contingency simulation, the Mean Absolute Error (MAE) between the results produced by COSMIC and PowerWorld was within 0.11\%. Since COSMIC adopted simplified exciter and governor models that are not included in many commercial packages, several of the time constants were set to zero or very close to zero in order to obtain agreement between the two models. 

\section{Experiments and Results} \label{sec:results}
In this section we present the results from several computational experiments, which illustrate this model and the types of insight that one can gain with dynamic cascading failure simulation. Three test systems are used: the 9-bus system \cite{Anderson:2008}, the 39-bus system, and the 2383-bus system \cite{Zimmerman:2011}, which is an equivalenced system based on the year 2000 winter snapshot for the Polish network. Section \ref{polar_rec} compares the computational efficiency of the polar and rectangular formulations of the model. Section \ref{relay_demo} uses the 9-bus test case to illustrate the different relay functions and their time delay algorithms. Section \ref{cascading_demo} demonstrates how COSMIC processes cascading events such as line branch outages, load shedding, and islanding. Section \ref{N2_polish} studies the impact of different load modeling assumptions on cascading failure sizes. Finally, Section \ref{cosmic_dc} compares results from COSMIC with results from a dc-power flow based model of cascading failure in order to understand similarities and differences between these two different modeling approaches.

\subsection{Polar formulation vs.~rectangular formulation in computational efficiency} \label{polar_rec}
COSMIC includes both polar and rectangular power flow formulations. In order to compare the computational efficiency of the two formulations, we conducted a number of $N-1$ and $N-2$ experiments using two different cases, the 39-bus and 2383-bus systems. The amount of time that a simulation requires and the number of linear solves ($\mathbf{A}x=b$) are two measures commonly used to evaluate the computational efficiency of a model. Compared to the first metric, the number of linear solves describes computational speed independently of specific computing hardware, and it is adopted in this study. 

Tables~\ref{table_1} and \ref{table_2} compare the two formulations with respect to the number of linear solves. These comparisons are grouped by the amount of demand lost in each simulation, given that longer cascades tend to require more linear solves and have different numerical properties. For the 39-bus case, 45 $N-1$ experiments and 222 randomly selected $N-2$ experiments were conducted, and each of them finished at 50 seconds and lost the same amount of power demand. As shown in Table~\ref{table_1}, the performances of the two methods were similar in terms of number of linear solves; however, the rectangular formulation required fewer linear solves and shows various improvements (e.g., positive decrease rectangular vs.~polar) for different demand losses.

For the 2383-bus test case, we simulated 2494 $N-1$ and 556 $N-2$ contingencies; Table~\ref{table_2} shows the results. There was no significant improvement for the rectangular formulation over the polar formulation, and the number of linear solves that resulted from both forms were almost identical.

One can also notice that solving the 2383-bus case required fewer linear solves than for the 39-bus case. This suggests that some branch outages have a higher impact on a smaller network and cause more dynamic oscillations than on a larger network such as the 2383-bus case.
\begin{table}
\centering
\scriptsize
\caption{The number of linear solves for the 39-bus case.}
\begin{tabular}{|c|c|c|c|c|c|c|}
\hline
\multirow{2}{*}{}& 0-1&1-200& 200-500 & 500-1000 & 1000-3000& 3000-6000\\
&MW & MW&  MW  & MW  & MW &MW \\
\hline
\textbf{Polar$^1$} & 3084& 6839 & 10523 & 5994 & 41067 & 10278\\
\hline
\textbf{Rec.$^1$} & 3035 & 6724 & 10324 & 5846 & 4052 & 9501\\
\hline
\hline
\textbf{Tests$^2$} & 176 & 38 & 14 & 32 & 4 & 3 \\
\hline
\textbf{\% Dec.$^3$} & 1.59\% & 1.69\% & 1.89\% & 2.46\% & 1.33\% & 7.6\% \\
\hline
\end{tabular}
\\ \footnotesize \textbf{1:} The density (non-zero rates) of the Jacobian matrices for polar and rectangular forms are 0.0376\% and 0.0383\% respectively; \textbf{2:} the number of tests; \textbf{3:} The percentage decrease in the number of linear solves, rectangular vs.~polar. \\ 
\vspace{0.05in}
\label{table_1}
\footnotesize
\caption{Comparison of linear solves for the 2383-bus case.}
\begin{tabular}{|c|c|c|}
\hline
& 0-1 MW&1-2000 MW\\
\hline
\textbf{Polar$^4$} & 867.77 & 692.81 \\
\hline
\textbf{Rec.$^4$} & 867.84 & 693.10 \\
\hline
\hline
\textbf{Tests} & 2494 & 556  \\
\hline
\textbf{\% Dec.} & -0.009\% & -0.04\% \\
\hline
\end{tabular}
\\ \footnotesize \textbf{4:} The density (non-zero rates) of the Jacobian matrices for polar and rectangular formulations are 0.000856\% and 0.000872\% respectively.
\label{table_2}
\end{table}

\subsection{Relay event illustration} \label{relay_demo}
\begin{figure}[h]
\centering
\includegraphics[width=1\columnwidth]{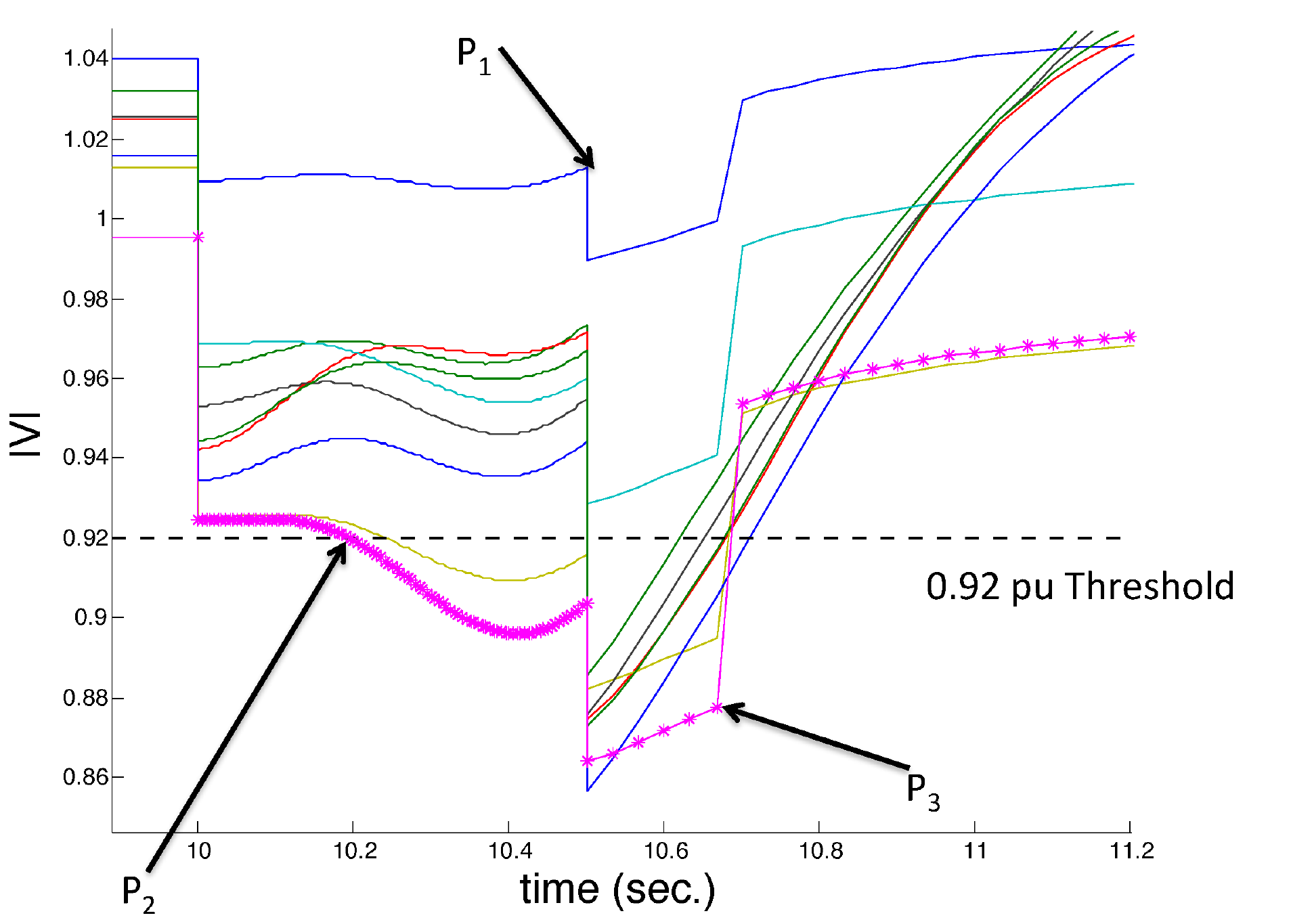}
\caption{The bus voltage magnitudes when the branch from Bus 6 to Bus 9 in the 9-bus system is tripped. DIST relay and UVLS relay actions are illustrated. The dashed black line is the the voltage threshold for UVLS relays.}
\label{Fig:Vmag_zoomed}
\end{figure}

To depict the functionality of how protective relays integrate with COSMIC's time delay features, we implemented the following example using the 9-bus system. The initial event was a single-line outage from Bus 6 to Bus 9 at $t=10$ seconds. The count-down timer of the DIST relay for branch 5-7 was activated with a $t_{\mbox{preset-delay}}=0.5$ seconds. As shown in Fig.~\ref{Fig:Vmag_zoomed}, the system underwent a transient swing following the one-line outage. Right after $0.5$ seconds, $t_{\mbox{delay}}$ ran out ($P_{1}$) and branch 5-7 was tripped by its DIST relay, which resulted in two isolated islands. Meanwhile, the thresholds for UVLS relays were set to $0.92$ pu. Note that the magenta voltage trace violated this voltage limit at about $t=10.2$ seconds ($P_{2}$). Because this trace continued under the limit after that, its UVLS timer counted down from $t_{\mbox{preset-delay}}=0.5$ seconds till $t=10.7$ seconds ($P_{3}$), where its UVLS relay took action and shed $25\%$ of the initial load at this bus. The adjacent yellow trace illustrates that the UVLS relay timer was activated as well but with a small lag. This UVLS relay never got triggered because before its $t_{\mbox{delay}}$ emptied out the load shedding at $P_{2}$ put this yellow trace back upon threshold, and its $t_{\mbox{delay}}$ was restored to $0.5$ seconds.

\subsection{Cascading outage examples using the 39-bus and the 2383-bus power systems} \label{cascading_demo}
The following experiment demonstrates a cascading outage example using the IEEE 39-bus case (see Table~\ref{table_3} for a summary of the sequential events). The system suffered a strong dynamic oscillation after the initial two exogenous events (branches 2-25 and 5-6). After approximately 55 seconds the first OC relay at branch 4-5 triggered. Because the monitored current kept up-crossing and down-crossing its limit, it delayed the relay triggering based on the time delay algorithms for the protection devices. Load shedding at two buses (Bus 7 and Bus 8) occurred around $t=55.06$ seconds, then another two branches (10-13 and 13-14) shut down after OC relay trips at $t=55.28$ seconds. These events separated the system into two islands. At $t=55.78$ seconds, two branches (3-4 and 17-18) were taken off the grid and this resulted in another island. The system eventually ended up with three isolated networks. However, one of them was not algebraically solvable due to a dramatic power imbalance, and it was declared as a blackout area.
\begin{figure}[h]
\centering
\includegraphics[width=1\columnwidth]{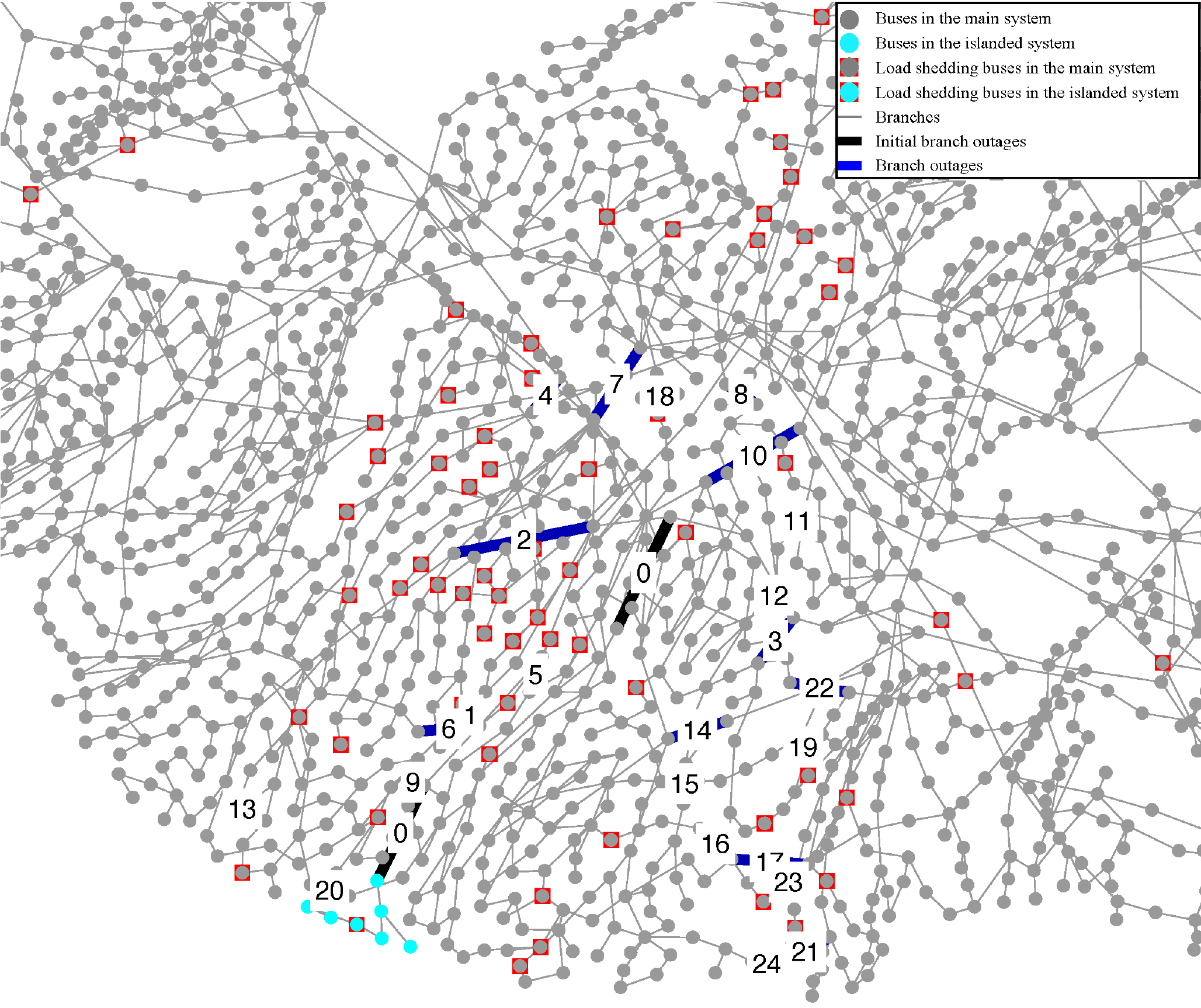}
\caption{The sequence of events for an illustrative cascading failure in the 2383-bus network. Numbers show the locations and sequence of line outages. Number 0 with black highlights denotes the two initial events. Other sequential numbers indicate the rest of the branch outages. In this example, 24 branches are off-line and causes a small island (the blue colored network) in the end. The dots with additional red square indicate buses where load shedding happens.}
\label{Fig:seq_polish}
\end{figure}
\begin{figure*}
\centering
\includegraphics[width=1\textwidth]{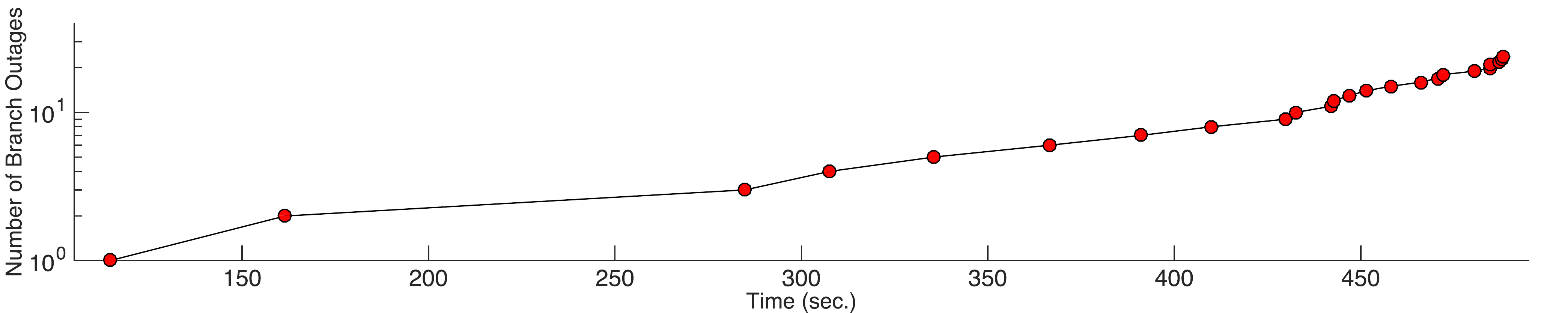}
\includegraphics[width=1\textwidth]{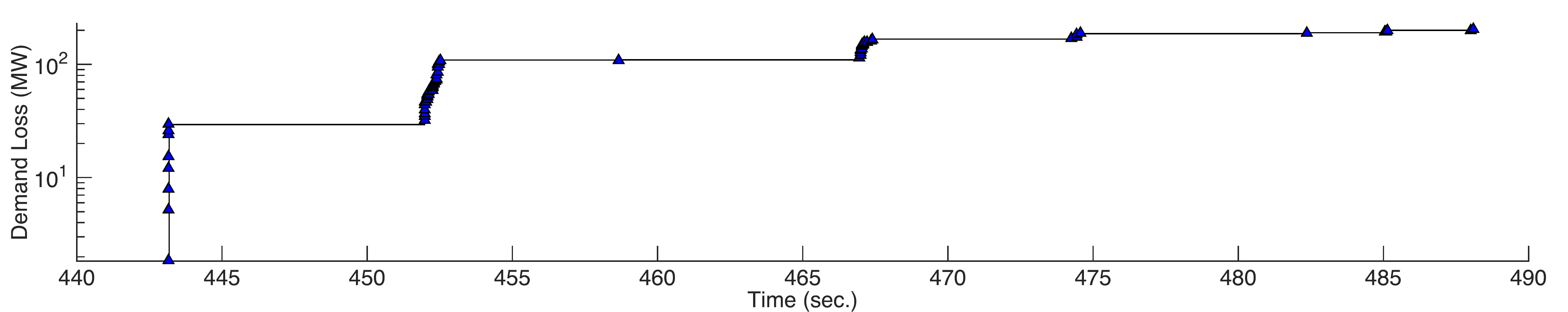}
\caption{The top panel shows the timeline of all branch outage events listed in Fig.~\ref {Fig:seq_polish}, and the lower panel zooms in the associated load-shedding events.}
\label{Fig:seq_polish_timeline}
\end{figure*}
\begin{table}[h]
\centering
\small
\caption{39-bus case cascading outage example.}
\begin{tabular}{|c|c|c|}
\hline
{\textbf{No.}}&{\textbf{Time (sec)}}& {\textbf{Events}}\\
\hline
1& \textbf{3} & Initial events: branches 2-25 and 5-6 trip \\
\hline
2& \textbf{54.78} & Branch 6-7 is tripped by OC relay\\
\hline
3& \textbf{55.06} & 58.45 MW load shedding at Bus 7 \\
\hline
4& \textbf{55.07} & 130.50 MW load shedding at Bus 8\\
\hline
5& \textbf{55.28} & Branch 4-14 is tripped by OC relay\\
\hline
6& \textbf{55.28} & Branch 10-13 is tripped by OC relay\\
\hline
7& \textbf{55.28} & Branch 13-14 is tripped by OC relay\\
\hline
--- & \textbf{55.28} & 1st islanding event\\
\hline
8& \textbf{55.78} & Branch 3-4 is tripped by OC relay\\
\hline
9& \textbf{55.78} & Branch 17-18 is tripped by OC relay\\
\hline
--- & \textbf{55.78} & 2nd islanding event\\
\hline
\end{tabular}
\label{table_3}
\end{table}

Fig.~\ref {Fig:seq_polish} illustrates a sequence of branch outage events using the 2383-bus network. This cascading was initiated by two exogenous branch outages (Branches 31-32 and 388-518, which are marked as number 0 in Fig.~\ref {Fig:seq_polish}) and resulted in a total of 92 discrete events. Out of these, 24 were branch outage events, and they are labeled in order in Fig.~\ref {Fig:seq_polish}. These events consequently caused a small island (light blue colored dots). The dots with additional red square highlighting indicate buses where load shedding occurred. From this figure we can see that cascading sequences do not follow an easily predictable pattern, and the affected buses with low voltage or frequency may be far from the initiating events. 

The top panel in Fig.~\ref {Fig:seq_polish_timeline} shows the timeline of all branch outage events for the above cascading scenario, and the lower panel zooms in the load-shedding events. In the early phase of this cascading outages, the occurrence of the components failed relatively slowly, however, it speeded up as the number of failures increased. Eventually the system condition was substantially compromised, which caused fast collapse and the majority of the branch outages as well as the load shedding events (see lower panel in Fig.~\ref {Fig:seq_polish_timeline}). 

\subsection{$N-2$ contingency analysis using the 2383-bus case} \label{N2} \label{N2_polish}
Power systems are operated to ensure the $N-1$ security criterion so that any single component failure will not cause subsequent contingencies \cite{Ren:2008}. The modified 2383-bus system that we are studying in this paper satisfies this criterion for transmission line outages. Thus, we assume here that branch outages capture a wide variety of exogenous contingencies that initiate cascades, for example a transformer tripping due to a generator failure. 

The experiment implemented here included four groups of 1200 randomly selected $N-2$ contingencies for the 2383-bus system. We measured the size of the resulting cascades using the number of relay events and the amount of demand lost. Each group had a different static load configuration. The load configuration for the first group was 100\% constant Z load; the second group used 100\% E load; the third had 100\% constant P load; and the fourth one included 25\% of each portion in the ZIPE model. We set TEMP, DIST, UVLS, and UFLS relays active in this experiment and deactivated OC relay due to its overlapping/similar function with TEMP relay.

\begin{figure}[h]
\centering
\includegraphics[width=1\columnwidth]{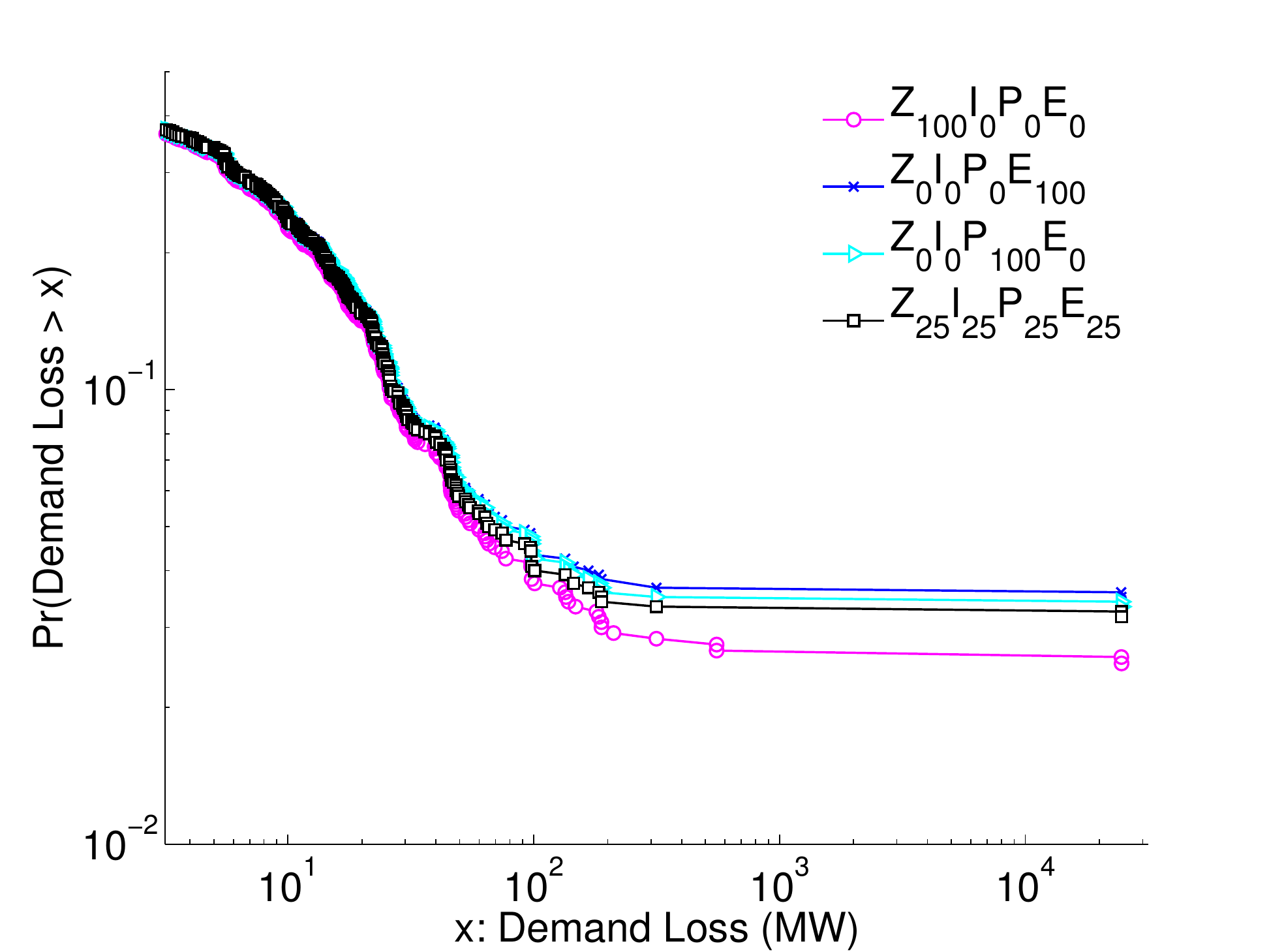}
\caption{CCDF of demand losses for 1200 randomly selected $N-2$ contingencies using the 2383-bus case. $Z_{100}I_0P_0E_0$ indicates 100\% constant Z load and 0\% of other load portions; similarly $Z_0I_0P_0E_{100}$, $Z_0I_0P_{100}E_0$ and $Z_{25}I_{25}P_{25}E_{25}$ represent constant E load, constant P load and a combination of equal amount of the four load types, respectively. }
\label{Fig:ccdf2383_dloss}
\end{figure}

Fig.~\ref{Fig:ccdf2383_dloss} shows the Complementary Cumulative Distribution Function (CCDF) of demand losses for these four groups of simulations. The CCDF plots of demand losses exhibit a heavy-tailed blackout size distribution, which are typically found in both historical blackout data and cascading failure models \cite{Hines:2009}. The magenta trace indicates constant Z load, and shows the best performance --in terms of the average power loss and the probability of large blackout-- within this set of 1200 random $N-2$ contingencies (listed in Table~\ref{table_4}). In contrast, the blue trace (constant E load) reveals the highest risk of large size blackouts ($>1000$ MW). The constant P load has a similar trend as the constant E load, due to their similar stiff characteristics; however, the constant E load with this particular exponent, 0.08, demonstrates a negative effect on the loss of load. The one with 25\% Z, 25\% I, 25\% P and 25\% of E performs in the middle of constant P load and constant Z load.  

As can be seen in Table~\ref{table_4}, the probabilities of large demand losses varies from 2.5\% to 3.5\% for those four load configurations. These results show that load models play an important role in dynamic simulation and may increase the frequency of non-convergence if they are not properly modeled. 
\begin{table}[h]
\centering
\footnotesize
\caption{The average demand loss, average branch outages and the probabilities of loss of the whole system for different load models.}
\begin{tabular}{|c|c|c|c|}
\hline
{\textbf{Load Model}}& {\textbf{Avg. Loss}} &  {\textbf{Avg. BO$^1$}} &  {\textbf{Prob. of Largest Loss}} \\
\hline
$Z_{100}$  & 644.19 MW& 0.2092 & 0.025\\
\hline
$E_{100}$  & 889.02 MW &  0.1800  & 0.035\\
\hline
$P_{100}$ & 848.21  MW &  0.1775  & 0.033\\
\hline
$Z_{25}I_{25}P_{25}E_{25}$  & 807.11 MW & 0.2042  & 0.032\\
\hline
\end{tabular}
\\ \footnotesize \textbf{1:} BO --- branch outages.
\label{table_4}
\end{table}

\begin{figure*}[th!]
\centering
\includegraphics[width=1\textwidth]{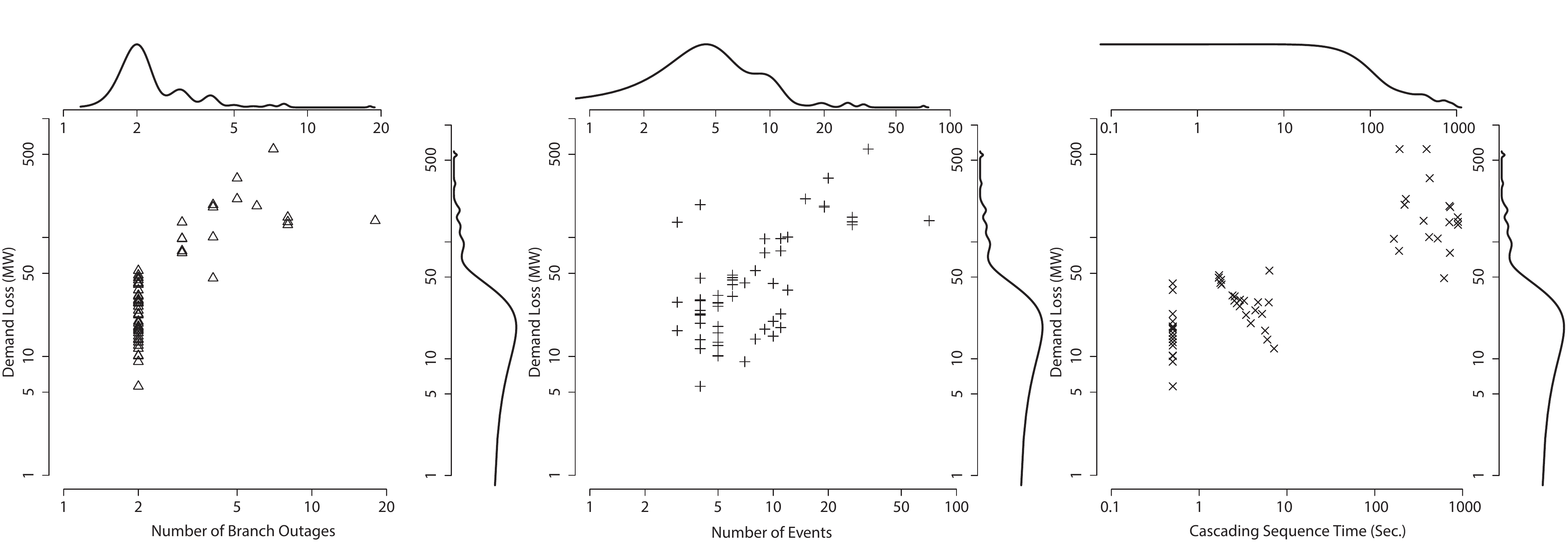}
\caption{Comparison of various measures of cascade size in the 88 $N-2$ contingency simulations that have at least one dependent event after the two initiating contingencies (31 of the 1200 simulations did not converge and were declared as full blackouts, and 1081 of them were resilient cases without any dependent events). The left panel compares the amount of load shedding to the number of branch outages, the middle panel compares demand losses to the total number of discrete events, and the right panel shows the correlation between the amount of time until the cascade stabilized with the amount of demand lost (the time between first and last events). The density curves of the $x$ and $y$ variables are shown on the top and right of each panel, respectively. Note that the event counts include the two initiating contingencies.}
\label{Fig:dloss_vs_others}
\end{figure*}

\begin{figure}[h]
\centering
\includegraphics[width=1\columnwidth]{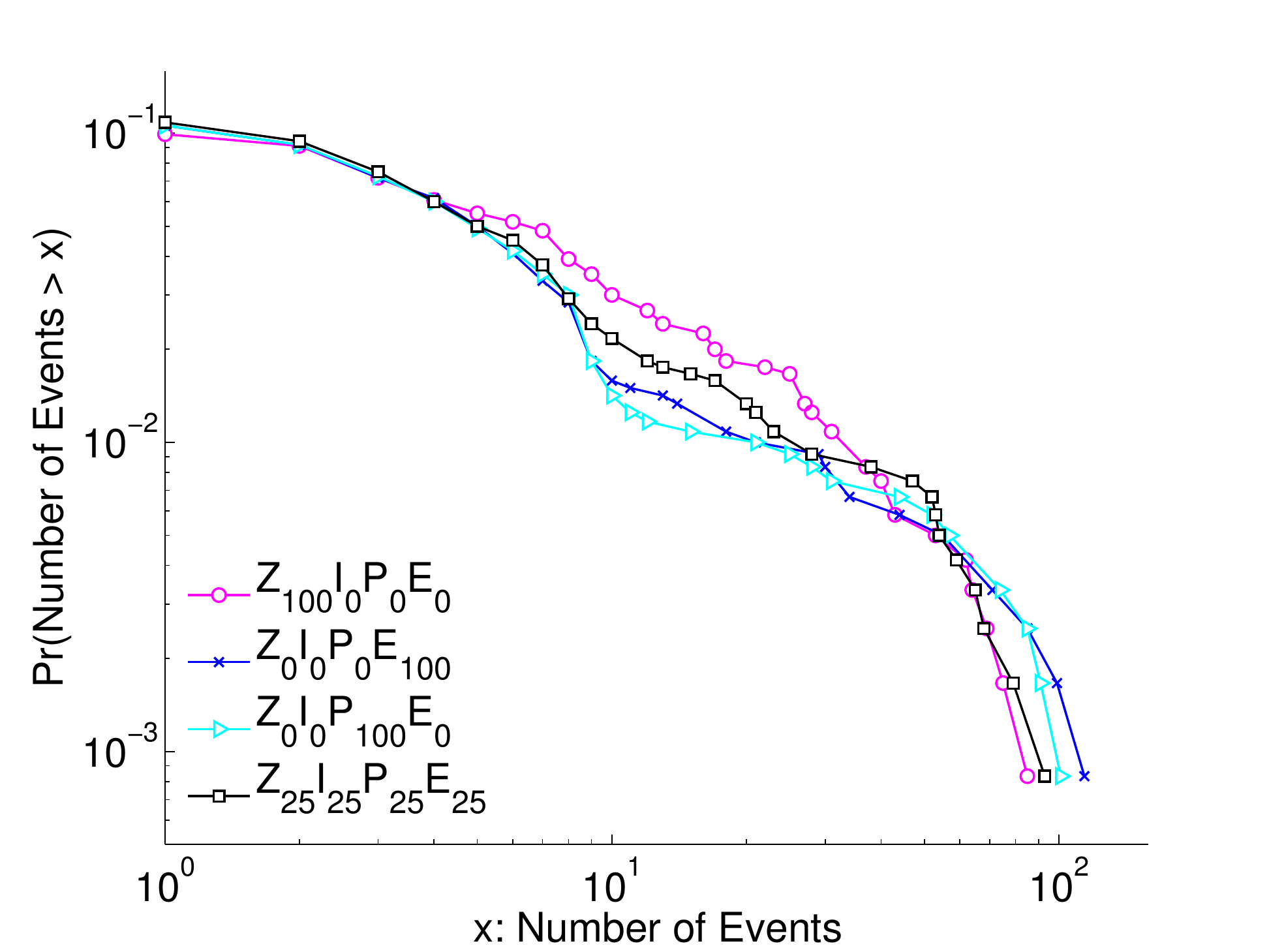}
\caption{CCDF of event length for 1200 randomly selected $N-2$ contingencies using the 2383-bus case.}
\label{Fig:ccdf2383_event_length}
\end{figure}

\begin{figure}[h]
\centering
\includegraphics[width=1\columnwidth]{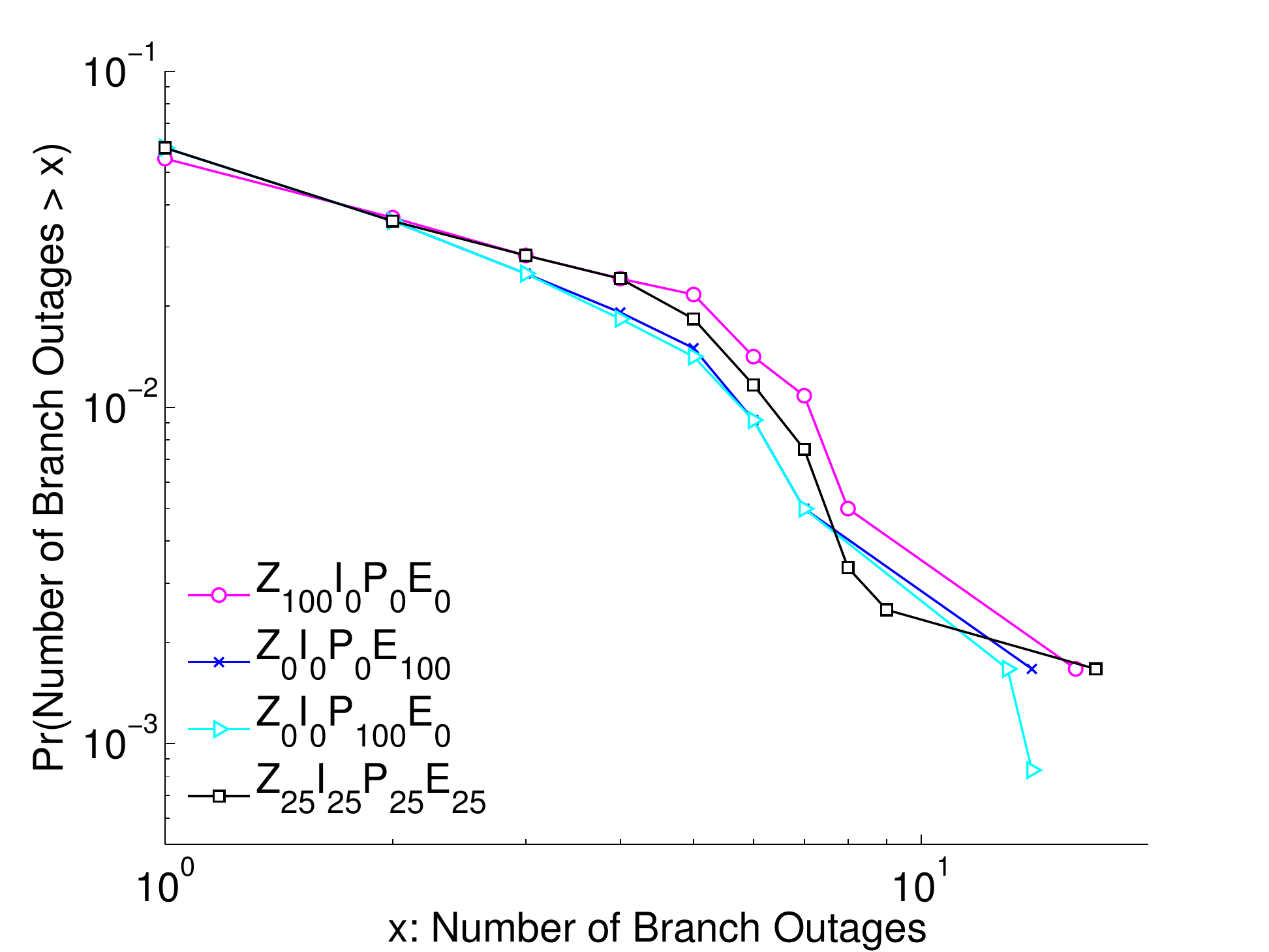}
\caption{CCDF of event length (branch outages only) for 1200 randomly selected $N-2$ contingencies using the 2383-bus case.}
\label{Fig:ccdf2383_bo_length}
\end{figure}

Fig.~\ref{Fig:ccdf2383_event_length} shows the CCDF plots of total dependent event length, including all event types after the initial contingencies, such as branch outages caused by TEMP and DIST relays, and load shedding events by UVLS and UFLS relays. Fig.~\ref{Fig:ccdf2383_bo_length} shows the CCDF of the branch outage lengths only. We can see from these two figures that the distributions of constant P and constant E loads have a comparable pattern, and they are in general less likely to have the same amount of branch outages, relative to the other two configurations. In Fig.~\ref{Fig:dloss_vs_others} we compare the amount of load shedding in the 88 simulated cascades to the number of discrete events (including the two exogenous line outages), the total number of events, and the cascading time. Each of these 88 samples includes at least one dependent event (31 of the 1200 simulations did not converge and were declared as full blackouts, and 1081 of them were resilient cases without any dependent events). The results show that there is a positive, but weak, correlation between the number of events and the blackout size in MW. We also find that there is a positive, but weak, correlation between cascading failure size in MW and the length of the cascade in seconds.

\subsection{Comparison with a dc cascading outage simulator} \label{cosmic_dc}
A number of authors have implemented quasi-steady state (QSS) models using the dc power flow equations to investigate cascading outages \cite{Eppstein:2012,Carreras:2002,Yan:2014}. The dc model includes numerous simplifications that are substantially different from the ``real'' system. However, the dc model is numerically stable, making it possible to produce results that can be statistically similar to data from real power systems \cite{Carreras:2013}. On the other hand, dynamic models, such as the one presented here, include many mechanisms of cascading that cannot be represented in the dc model. In order to understand the implications of these differences, we conducted two experiments to compare COSMIC with the dc QSS model described in \cite{Eppstein:2012} with respect to the overall probabilities of demand losses and the extent to which the patterns of cascading from the two models agree. These experiments provide helpful insights into the dominant cascading mechanisms in different phases of cascading sequences, and into the conditions under which one would want to use a dynamic model as opposed to a simpler model.

\subsubsection{The probabilities of demand losses} 
\begin{figure}[h]
\centering
\includegraphics[width=1\columnwidth]{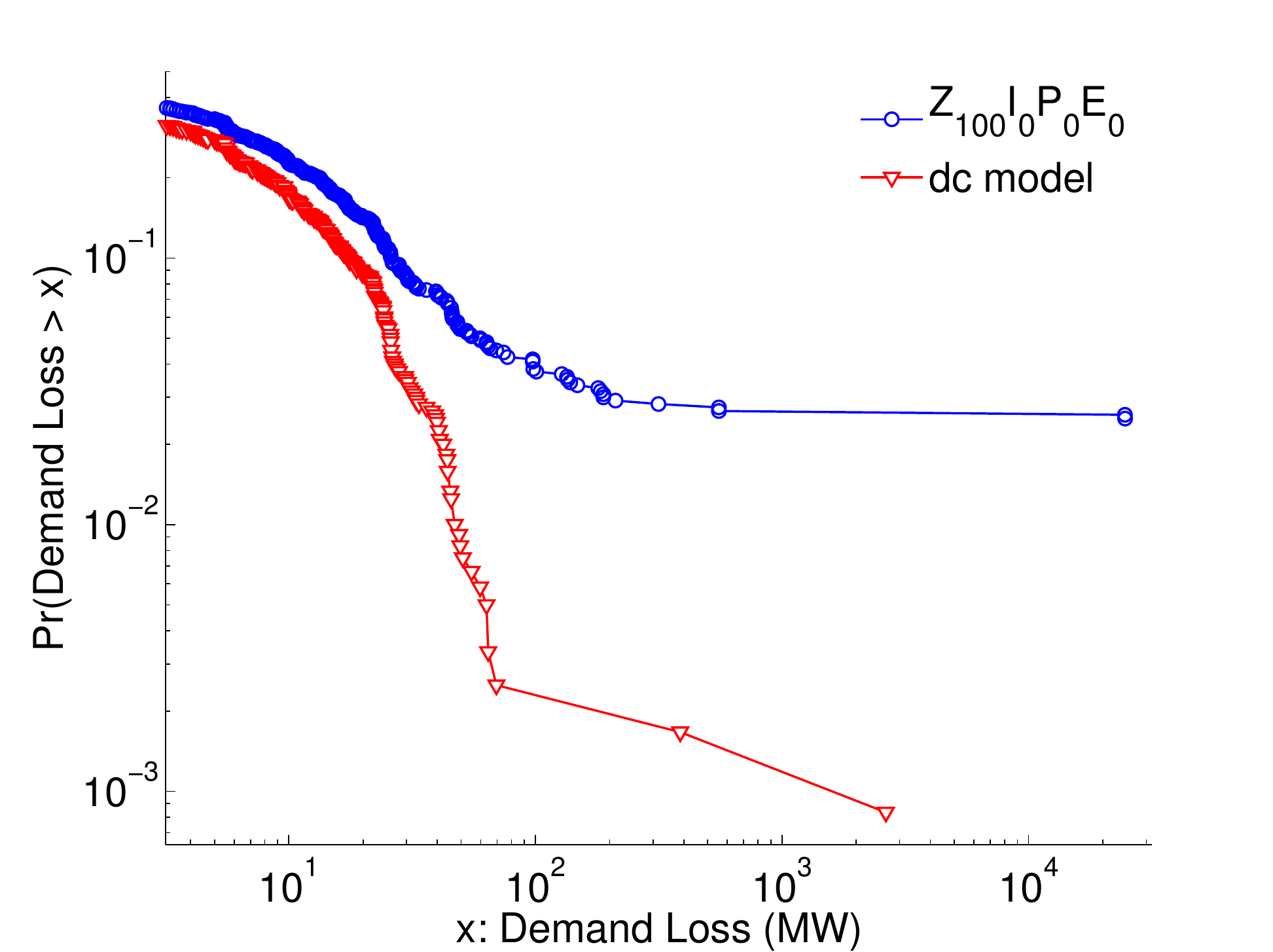}
\caption{CCDF of demand losses for COSMIC with constant impedance load and the dc cascading outage simulator, for 1200 randomly selected $N-2$ contingencies using the 2383-bus case.}
\label{Fig:ccdf_z_vs_dc}
\end{figure}

For the first experiment, we computed the CCDF of demand losses in both COSMIC (with the constant impedance load model) and the dc model using the same 1200 branch outage pairs from \ref{N2}. From Fig.~\ref{Fig:ccdf_z_vs_dc} one can learn that the probability of demand losses in the dc simulator is lower than that of COSMIC for the same amount of demand losses. In particular, the largest demand loss in dc simulator is much smaller than in COSMIC (2639 MW vs.~24602 MW, with probabilities 0.08\% vs.~2.5\%). This large difference between them is not surprising because the dc model is much more stable and does not run into problems of numerical non-convergence. Also, the protection algorithms differ somewhat between the two models. 
In addition, some of the contingencies do produce large blackouts in the dc simulator, which causes the fat tail that can be seen in Fig.~\ref{Fig:ccdf_z_vs_dc}.

Numerical failures in solving the DAE system greatly contributed to the larger blackout sizes observed in COSMIC, because COSMIC assumes that the network or sub-network in which the numerical failure occurred experienced a complete blackout. This illustrates a tradeoff that comes with using detailed non-linear dynamic models: while the component models are more accurate, the many assumptions that are needed substantially impact the outcomes, potentially in ways that are not fully accurate.

It is possible that this result may be affected to some extent by the size of the sample set. The 1200 randomly selected contingency pairs represent 0.0278\% of the total $N-2$ branch outage pairs. Extensive investigation of how different sampling approaches might impact the observed statistics remains for future work.

\begin{table}
\centering
\small
\caption{The statistical results for the comparison between COSMIC and dc simulator.}
\begin{tabular}{|c|c|c|}
\hline 
\multirow{2}{*}{} & \multicolumn{2}{c|}{\textbf{336 critical pairs}} \tabularnewline
\cline{2-3} 
 & \textbf{COSMIC} & \textbf{dc simulator}\tabularnewline
\hline 
\textbf{Avg. Demand Loss} & 23873 MW & 3039.5 MW \tabularnewline
\hline 
\textbf{Avg. Branch Outage Length} & 7.5476 & 26.7351 \tabularnewline
\hline 
\textbf{Avg. R} & \multicolumn{2}{c|}{0.1948} \tabularnewline
\hline 
\textbf{Std. Dev. of R} & \multicolumn{2}{c|}{0.1381} \tabularnewline
\hline 
\textbf{Max R} & \multicolumn{2}{c|}{0.6} \tabularnewline
\hline 
\textbf{Avg. R (first 10)$^1$} & \multicolumn{2}{c|}{0.3487} \tabularnewline
\hline 
\textbf{Std. Dev. of R (first 10)} & \multicolumn{2}{c|}{0.2445} \tabularnewline
\hline 
\textbf{Max R (first 10)} & \multicolumn{2}{c|}{1} \tabularnewline
\hline 
\end{tabular}
\\ \footnotesize \textbf{1: } only the first 10 branch outages.
\label{table_5}
\end{table}
   
\subsubsection{Path Agreement Measurement} 
The second comparison experiment was to study the patterns of cascading between these two models. This comparison provides additional insight into the impact of dynamics on cascade propagation patterns. In order to do so, we compared the sets of transmission lines that failed using the ``Path Agreement Measure" introduced in \eqref{Eq:path} \cite{Fitzmaurice:2012}. The relative agreement of cascade paths, $R(m_1 ,m_2)$, is defined as follows. If models $m_1$ and $m_2$ are both subjected to the same set of exogenous contingencies: $C = \{c_1,c_2,c_3,.....\}$. $R(m_1 ,m_2)$ measures the average agreement in the set of dependent events that result from each contingency in each model. If contingency $c_i$ results in the set $A_i$ of dependent branch failures in model $m_1$ and the set $B_i$ of dependent branch failures in $m_2$, $R(m_1 ,m_2)$ is defined as: 
\begin{equation}
R(m_{1},m_{2})=\frac{1}{C}\overset{|C|}{\underset{i=1}{\sum}}\frac{|A_{i}\cap B_{i}|}{|A_{i}\cup B_{i}|}
\label{Eq:path}
\end{equation}

The experiment measured $R(m_1 ,m_2)$ between COSMIC (with a $Z_{25}I_{25}P_{25}E_{25}$ load configuration) and the dc simulator for 336 critical branch outage pairs. These branch outage pairs are the ones reported in~\cite{Eppstein:2012} to produce large cascading blackouts. Table~\ref{table_5} shows that the average $R$ between the two models for the whole set of sequences is $0.1948$, which is relatively low. This indicates that there are substantial differences between cascade paths in the two models. Part of the reason is that the dc model tends to produce longer cascades and consequently increase the denominator in \eqref{Eq:path}. In order to control this, we computed $R$ only for the first 10 branch outage events. The average R increased to 0.3487, and some of the cascading paths showed a perfect match ($R=1$). This suggests that the cascading paths resulting from the two models tend to agree during the early stages of cascading, when non-linear dynamics are less pronounced, but disagree during later stages.

\section{Conclusions} \label{sec:conclusion}
This paper describes a method for and results from simulating cascading failures in power systems using full non-linear dynamic models. The new model, COSMIC, represents a power system as a set of hybrid discrete/continuous differential algebraic equations, simultaneously simulating protection systems and machine dynamics. Several experiments illustrated the various components of COSMIC and provided important general insights regarding the modeling of cascading failure in power systems.
By simulating 1200 randomly chosen $N-2$ contingencies for a 2383-bus test case, we found that COSMIC produces heavy-tailed blackout size distributions, which are typically found in both historical blackout data and cascading failure models \cite{Hines:2009}. However, the relative frequency of very large events may be exaggerated in dynamic model due to numerical non-convergence (about $3\%$ of cases). More importantly, the blackout size results show that load models can substantially impact cascade sizes---cases that used constant impedance loads showed consistently smaller blackouts, relative to constant current, power or exponential models. In addition, the contingency simulation results from COSMIC were compared to corresponding simulations from a dc power flow based quasi-steady-state cascading failure simulator, using a new metric. The two models largely agreed for the initial periods of cascading (for about 10 events), then diverged for later stages where dynamic phenomena drive the sequence of events.

Together these results illustrate that detailed dynamic models of cascading failure can be useful in understanding the relative importance of various features of these models. The particular model used in this paper, COSMIC, is likely too slow for many large-scale statistical analyses, but comparing detailed models to simpler ones can be helpful in understanding the relative importance of various modeling assumptions that are necessary to understand complicated phenomena such as cascading.

\section{Appendix}  \label{sec:appendix}
Section \ref{app_A} presents the detailed dynamic component models in COSMIC, to further explain the dynamic equations (in Section \ref{sec:dae}). Section \ref{app_B} explains the formulation of the numerical solver in this model to solve the continuous integration. Section \ref{app_C} shows the relevant settings for the numerical solver and protective relays in COSMIC. This is free open source code with a GNU General Public License and is available for download and contributions in a web repository \cite{COSMIC:2015}.
\subsection{Differential equations in dynamic power system} \label{app_A}
The following differential equations are used to represent machine dynamics in COSMIC, based on the polar formulation.
\subsubsection{Equation for rotor speed --- Swing Equation}
The rotor speed related to bus $i$ is represented in the standard second order swing equation \cite{Kundur:1994}, which describes the rotational dynamics of a synchronous machine. The normalized rotor speed $\omega$ is fixed during the normal operation, and will accelerate or decelerate when there's disturbance. 
\begin{equation}
\label{eqs:swing}
M\frac{d{\omega_{i}}}{dt} = P_{m,i}-P_{g,i}-D\left({\omega_{i}}-1\right)
\end{equation}
where $\forall i\in N_{G}$ ($N_G$ is the set of all generator buses), $M$ is a machine inertia constant, $D$ is a damping constant, $P_{m,i}$ is the mechanical power input, and $P_{g,i}$ is the generator power output.

\subsubsection{Equation for rotor angle}
The rotor angle, $\delta_i$, is the integral of the relative rotor speed change with respect to synchronous speed (1.0 in per unit notation). It is given by the equation:
\begin{equation}
\label{eqs:ddelta_dt}
\frac{d\delta_{i}(t)}{dt}=2\pi f_{0}\left({\omega_{i}}-1\right)
\end{equation}

\subsubsection{Generator}
The salient-pole model is adapted in COSMIC. The active and reactive power outputs are given by the nonlinear equations \cite{Bergen:2009}: 
\begin{equation}
P_{g,i}=\frac{\left|E_{a,i}'\right|\left|V_{i}\right|}{X_{d,i}'}\sin\delta_{m,i}+\frac{\left|V_{i}\right|^{2}}{2}\left(\frac{1}{X_{q,i}}-\frac{1}{X_{d,i}'}\right)\sin2\delta_{m,i} 
\end{equation}
\begin{equation}
Q_{g,i}=\frac{\left|E_{a,i}'\right|\left|V_{i}\right|}{X_{d,i}'}\cos\delta_{m,i}+\left|V_{i}\right|^{2}\left(\frac{\cos^{2}\delta_{m,i}}{X_{d,i}'}+\frac{\sin^{2}\delta_{m,i}}{X_{q,i}}\right)
\label{eqs:power_salient}
\end{equation}
where $\forall i\in N_{G}$, $X_{d,i}$ and $X_{d,i}^{'}$ are the direct axis generator synchronous and transient reactances, respectively. The transient open circuit voltage magnitude \cite{Bergen:2009}, $|E_{a,i}^{'}(t)|$, is determined by the differential equation:
\begin{multline}
\frac{d\left|E_{a,i}^{'}(t)\right|}{dt}=-\left|E_{a,i}^{'}\right|\frac{X_{d,i}}{T_{\text{do},i}^{'}X_{d,i}^{'}}+\\\left(\frac{X_{d,i}}{X_{d,i}^{'}}-1\right)\frac{\left|V_{i}(t)\right|}{T_{\text{do},i}^{'}}\cos\left(\delta_{m,i}(t)\right)+\frac{E_{\text{fd},i}}{T_{\text{do},i}^{'}}
\label{eqs:open_circuit_voltage}
\end{multline}
where $T_{\text{do},i}$ is the direct axis transient time constant, and $E_{\text{fd},i}$ is the machine exciter output. This equation, along with Eqs.~\ref{eqs:swing}-\ref{eqs:ddelta_dt}, describe the basic physical properties of the generation machine, which results in a third order differential equation system. 

\subsubsection{Exciter}
\begin{figure}[h]
\centering
\includegraphics[width=1.0\columnwidth]{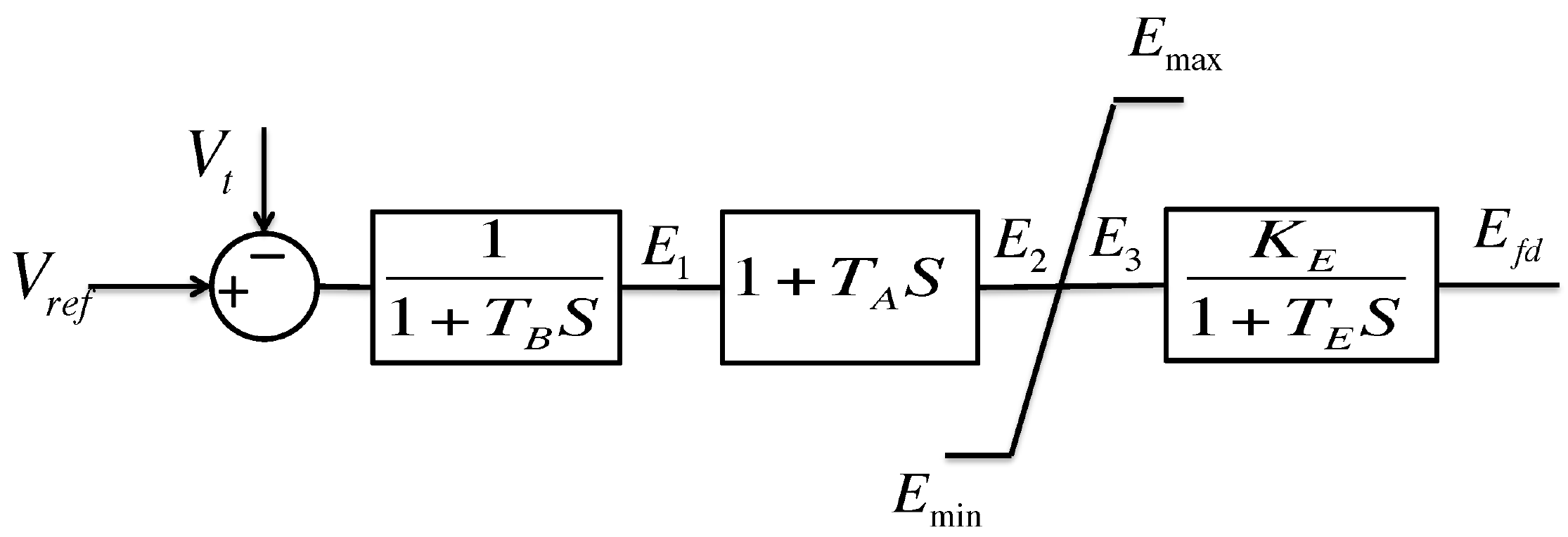}
\caption{The generic exciter model used in COSMIC}
\label{Fig:exc}
\end{figure}

The machine exciter control in COSMIC utilizes a generic model, which is a second order differential system:
\begin{equation}
\begin{split}
\frac{d\left|E_{\text{fd}}\right|}{dt}=\frac{1}{T_{E}} \left\{ K_{E}\cdot \text{sigm} \left[ \left(1-\frac{T_{A}}{T_{B}}\right) E_{1}+ \right. \right. \\ 
\left. \left. \frac{T_{A}}{T_{B}}\left(V_{\text{ref}}-V_{t}\right)\right]-E_{\text{fd}} \right\} 
\end{split}
\label{eq:Efd} 
\end{equation}
\begin{equation}
\begin{split}\frac{d\left|E_{1}\right|}{dt}=\frac{1}{T_{B}}\left(V_\text{ref}-V_{t}-E_{1}\right)\end{split}
\label{eq:E1}
\end{equation}
where $V_\text{ref}$ is the desired reference voltage, $V_t$ is the actual terminal voltage, $T_{A}$, $T_{B}$ and $K_{E}$ are the exciter time constants, and \text{sigm$(\cdot)$} is a differentiable sigmoidal function that acts as a limiter between $E_\text{min}$ and $E_\text{max}$. Fig.~\ref{Fig:exc} illustrates the simplified exciter configuration. 

\begin{figure}[h]
\centering
\includegraphics[width=1.0\columnwidth]{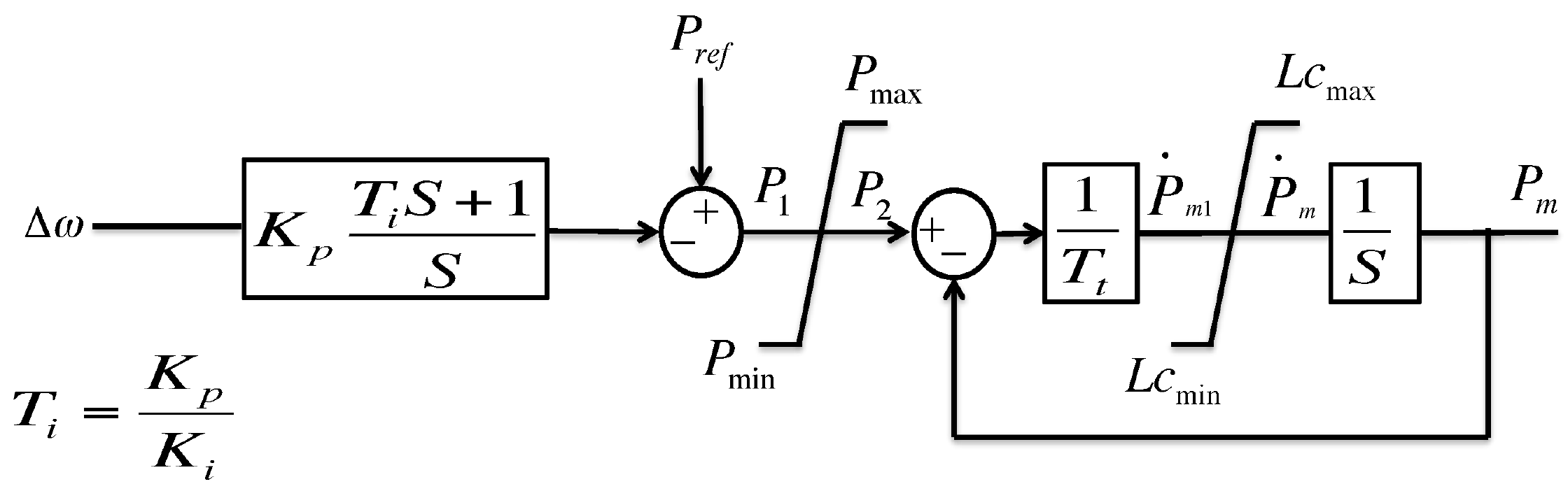}
\caption{Governor model used in COSMIC}
\label{Fig:gov}
\end{figure}

\subsubsection{Governor}
The machine governor controls the mechanical forcing given a deviation in the machine speed. COSMIC adapts a governor model that includes both proportional (droop) control and an approximate representation of secondary (integral) control in order to ensure that the system can restore frequency deviations to zero. It can be described by Eqs.~\eqref{eq:Pm}-\eqref{eq:P3}:
\begin{equation}
\frac{dP_{m}}{dt}=\text{sigm}\left\{\frac{1}{T_{t}}\left[\text{sigm}\left(P_\text{ref}-\left(\frac{1}{R}\Delta\omega+P_{3}\right)\right)-P_{m}\right]\right\}
\label{eq:Pm}
\end{equation}
\begin{equation}
\begin{split}\frac{dP_{3}}{dt}=\frac{1}{R\cdot T_{i}}\Delta\omega \end{split} 
\label{eq:P3}
\end{equation}
where $R$, $T_{i}$  and $T_{t}$ are the droop, PI time constants and servo-motor time constants, respectively. $P_{3}$ is a intermediate differential variable in the PI controller. Fig.~\ref{Fig:gov} illustrates the interactions among the governor variables.

\subsection{Trapezoidal method to solve DAE system} \label{app_B}
Let $t$ be the current time step, and $t+\Delta t$ be the next time step. Assuming that we have a consistent set of variables $\mathbf{x=}\mathbf{x}(t),\mathbf{y=}\mathbf{y}(t)$,and $\mathbf{z=}\mathbf{z}(t)$ for the current time $t$, $\mathbf{f}(t)$ and $\mathbf{g}(t)$ can be calculated from them. The trapezoidal solution method solves the following non-linear system to obtain $\mathbf{x}_{+}=\mathbf{x}(t+\Delta t)$ and $\mathbf{y_{+}}=\mathbf{y}(t+\Delta t)$: 
\begin{equation}
\mathbf{x}_{+} = \mathbf{x}+\frac{\Delta t}{2}\left[\mathbf{f}(t)+\mathbf{f}(t_{+},\mathbf{x}_{+},\mathbf{y}_{+},\mathbf{z})\right]
\label{eqs:trap-1}
\end{equation}
\begin{equation}
0 =  \mathbf{g}(t_{+},\mathbf{x}_{+},\mathbf{y}_{+},\mathbf{z})
\label{eqs:trap-2}
\end{equation}

\subsection{The relevant settings in COSMIC for this study} \label{app_C}
Table \ref{table_num_para} and Table \ref{table_relay_para} list the parameters that are used in the numerical solution, as well as the protective relay settings in COSMIC, respectively.  
\begin{table}[h]
\centering
\footnotesize
\caption{The parameters for the numerical solver in COSMIC}
\begin{tabular}{c|cc}
\hline
\textbf{Item}& \textbf{Setting 1} & \textbf{Setting 2} \\
\hline
\textbf{Convergence Tolerance} & $10^{-9}$ &---\\
\textbf{Maximum Number of Iterations} & 20 & ---\\
\textbf{Settings$^1$  to Increase $\Delta t$}& $2\cdot \Delta t_{\text{prev}}$ & 0.01 [mismatch]\\
\textbf{Settings$^2$ to Decrease $\Delta t$} & $1/2\cdot \Delta t_{\text{prev}}$ & 0.05 [mismatch]\\
\textbf{Maximum $\Delta t$} & 1 sec.& ---\\
\textbf{Minimum $\Delta t$} & 0.005 sec.& ---\\
\hline
\end{tabular}
\\ \footnotesize 1: If $\text{Max}(\text{mismatch})<0.01$, increase $\Delta t_{\text{next}}$ to $2\cdot \Delta t_{\text{prev}}$, except when a discrete event occurs. 2: If $\text{Max}(\text{mismatch})>0.05$, reduce $\Delta t_{\text{next}}$ to $1/2\cdot\Delta t_{\text{prev}}$, except when a discrete event occurs. \\ 
\label{table_num_para}
\end{table}
\begin{table*}[ht]
\centering
\small
\caption{The parameters for the relay settings in COSMIC}
\begin{tabular}{c|ccc}
\hline
& \textbf{Activation Threshold} &  \textbf{Delay} & \textbf{Other}\\
\hline
\textbf{OC relays} & rate-B / $\text{MVA}_{\text{base}}$&Time-inverse delay  &  $\left(\text{rate-B} / \text{MVA}_{\text{base}}\right)\cdot 150\%\cdot5$ sec.$^3$\\
\textbf{DIST relays} & 90\% of the line impedance & 0.5 sec. & --- \\ 
\textbf{TEMP relays} & temperature converted from rate-B & 0 sec.  & ---  \\ 
\textbf{UVLS relays} & 0.87 pu & 0.5 sec. &  25\%$^4$\\
\textbf{UFLS relays} & 0.95 pu & 0.5 sec. &  25\%$^4$\\
\hline
\end{tabular}
\\ \footnotesize 3: The triggering threshold for the OC relays. 4: Load shedding factor, indicating the percentage of shed power over total load for each load shedding action.\\ 
\label{table_relay_para}
\end{table*}

\bibliographystyle{IEEEtran}
\bibliography{cosmic_biblio}
\vspace{-0.1in}
\begin{IEEEbiography}[{\includegraphics[width=1in,height=1.25in,clip,keepaspectratio]{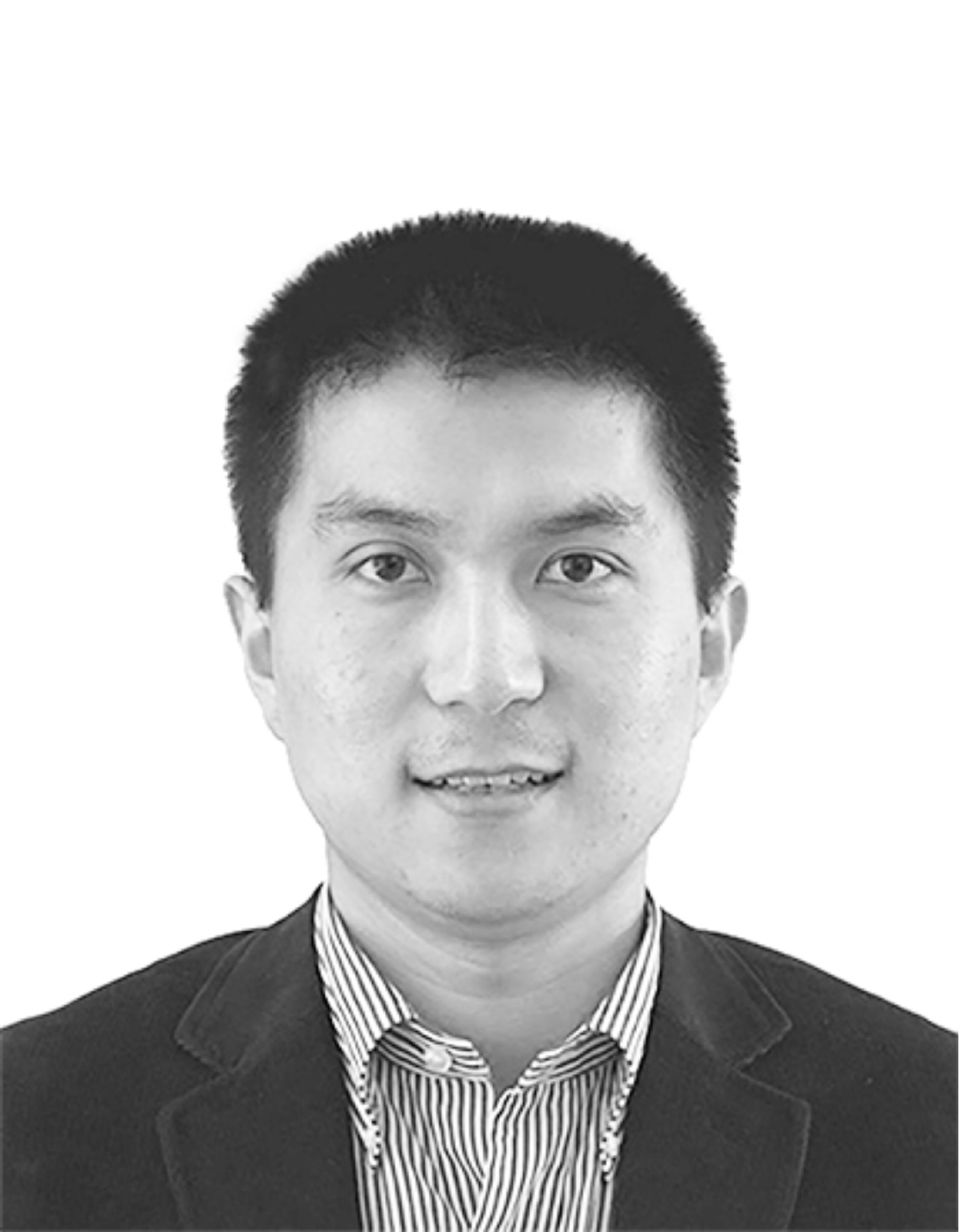}}]{Jiajia Song} (S`12) is currently pursuing his doctoral degree in Electrical Engineering at Oregon State University.

His research interests focus on dynamic power system and protection modeling, cascading outages analysis, and phasor measurement unit applications.     
\end{IEEEbiography}
\begin{IEEEbiography}[{\includegraphics[width=1in,height=1.25in,clip,keepaspectratio]{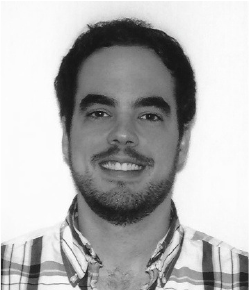}}]{Eduardo Cotilla-Sanchez} (S`08, M`12) received the M.S.~and Ph.D.~degrees in Electrical Engineering from the University of Vermont in 2009 and 2012, respectively.

He is currently an Assistant Professor in the School of Electrical Engineering \& Computer Science at Oregon State University. His primary field of research is the vulnerability of electrical infrastructure, in particular, the study of cascading outages. Cotilla-Sanchez is the Secretary of the IEEE Working Group on Understanding, Prediction, Mitigation and Restoration of Cascading Failures.
\end{IEEEbiography}
\begin{IEEEbiography}[{\includegraphics[width=1in,height=1.25in,clip,keepaspectratio]{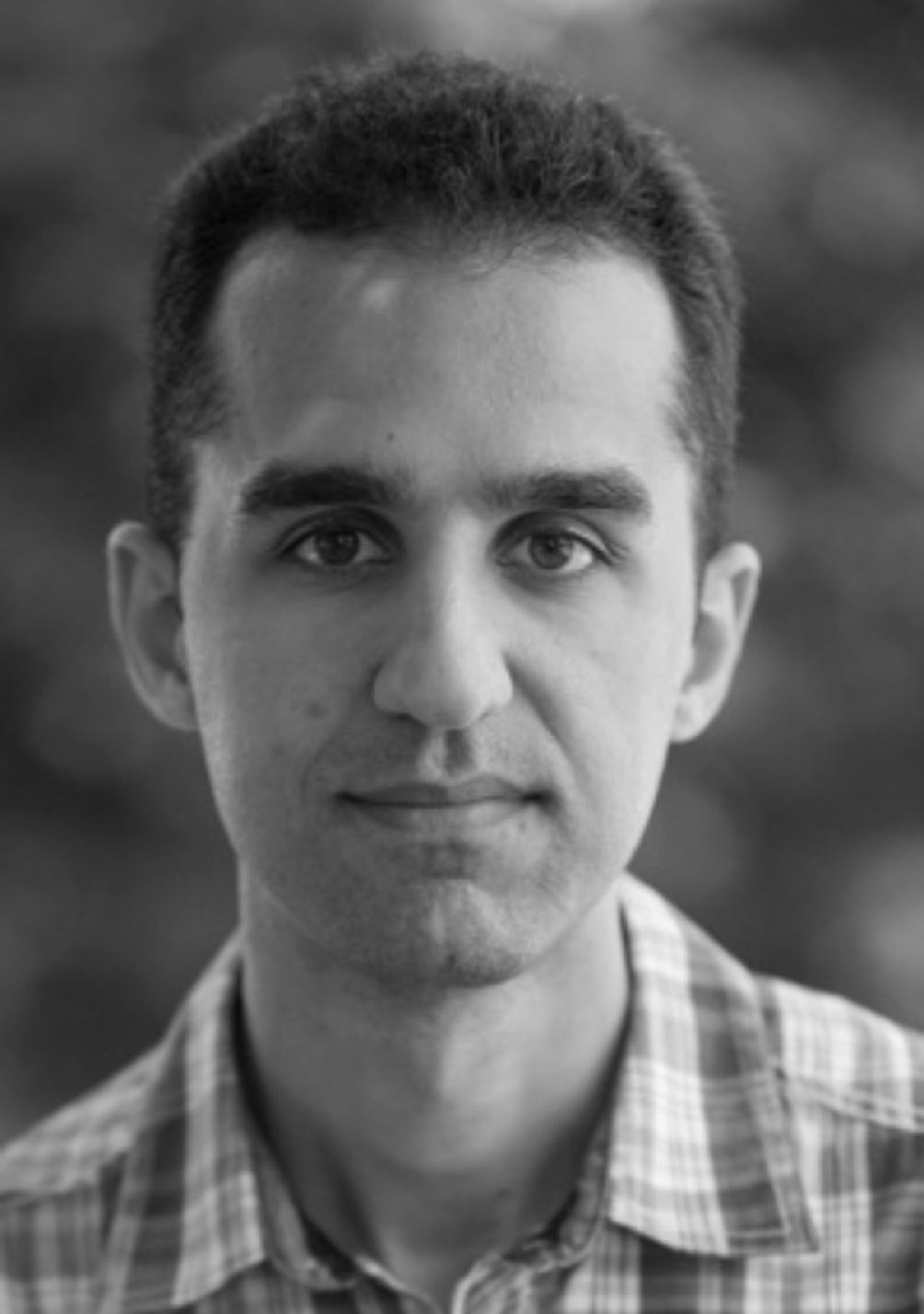}}]{Goodarz Ghanavati} (S`11) received the B.S. and M.S. degrees in electrical engineering from Amirkabir University of Technology, Tehran, Iran, in 2005 and 2008, respectively. Currently, he is pursuing the Ph.D. degree in Electrical Engineering at the University of Vermont, Burlington, VT, USA.
He worked as a Design Engineer at Monenco Iran Co. His research interests include power system dynamics, stochastic modeling of power systems, phasor measurement unit applications and smart grid.
\end{IEEEbiography}
\begin{IEEEbiography}[{\includegraphics[width=1in,height=1.25in,clip,keepaspectratio]{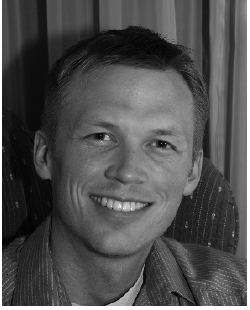}}]{Paul Hines} (S`96, M`07) received the Ph.D.~in Engineering and Public Policy from Carnegie Mellon University in 2007 and M.S.~(2001) and B.S.~(1997) degrees in Electrical Engineering from the University of Washington and Seattle Pacific University, respectively.

He is currently an Associate Professor in the School of Engineering at the University of Vermont, and a member of the adjunct research faculty at the Carnegie Mellon Electricity Industry center. Currently Dr. Hines serves as the vice-chair of the IEEE Working Group on Understanding, Prediction, Mitigation and Restoration of Cascading Failures.
\end{IEEEbiography}
\end{document}